\documentclass[preprint,review,12pt]{elsarticle}

\usepackage{amssymb}
\usepackage{amsmath,amsthm,amssymb,amsfonts}
\usepackage{algorithm}
\usepackage{algpseudocode}

\usepackage{booktabs}
\usepackage{hyperref}      

\journal{Heliyon}

\begin{document}

\begin{frontmatter}

\title{A General Description of Criticality in Neural Network Models}
\author[istbi]{Longbin Zeng}
\author[istbi,cni]{Jianfeng Feng}
\author[sms]{Wenlian Lu\corref{corr1}}
\cortext[cor1]{wenlian@fudan.edu.cn}
\affiliation[istbi]{organization={Institute of Science and Technology for Brain-Inspired Intelligence, Fudan University},
            city={Shanghai},
            country={China}}

\affiliation[sms]{organization={Shanghai Center for Mathematical Sciences, Fudan University},
            city={Shanghai},
            country={China}}
\affiliation[cni]{organization={Key Laboratory of Computational Neuroscience and BrainInspired Intelligence (Fudan University), Ministry of Education},
            state={China}}

\begin{abstract}
Recent experimental observations have supported the hypothesis that the cerebral cortex operates in a dynamical regime near criticality, where the neuronal network exhibits a mixture of ordered and disordered patterns. However, A comprehensive study of how criticality emerges and how to reproduce it is still lacking. In this study, we investigate coupled networks with conductance-based neurons and illustrate the co-existence of different spiking patterns, including asynchronous irregular (AI) firing and synchronous regular (SR) state, along with a scale-invariant neuronal avalanche phenomenon (criticality). We show that fast-acting synaptic coupling can evoke neuronal avalanches in the mean-dominated regime but has little effect in the fluctuation-dominated regime. In a narrow region of parameter space, the network exhibits avalanche dynamics with power-law avalanche size and duration distributions. We conclude that three stages which may be responsible for reproducing the synchronized bursting: mean-dominated subthreshold dynamics, fast-initiating a spike event, and time-delayed inhibitory cancellation. Remarkably, we illustrate the mechanisms underlying critical avalanches in the presence of noise, which can be explained as a stochastic crossing state around the Hopf bifurcation under the mean-dominated regime. Moreover, we apply the ensemble Kalman filter to determine and track effective connections for the neuronal network. The method is validated on noisy synthetic BOLD signals and could exactly reproduce the corresponding critical network activity. Our results provide a special perspective to understand and model the criticality, which can be useful for large-scale modeling and computation of brain dynamics.
\end{abstract}


\end{frontmatter}

\section{Introduction}
Criticality, usually characterized as neuronal avalanches, has been widely observed in both in vitro and in vivo biophysical experiments~\cite{friedman2012universal, plenz2007organizing, beggs2003neuronal}. In the critical state, population activities tend to operate at the boundary between order and disorder, occasionally and unpredictably appearing synchronized or irregularly spiking. The term “neuronal avalanches” describes a spiking pattern, in which the observed bursts of suprathreshold activity were interspersed by silence and showed a clear separation of time scales (their duration being much shorter than the inter avalanche intervals). The spatial and temporal distributions of neuronal avalanches have been identified following power-law statistics, implying that this brain state operates near a nonequilibrium critical point. Furthermore, previous studies have also shown that the whole-brain activity dynamics measured with noninvasive techniques, such as electroencephalography (EEG) and functional magnetic resonance imaging (fMRI), can also be well described by power-law statistics~\cite{yang2017co}.

\par
Until now, most studies have focused on scale-free behavior, showing a power law distribution of empirically observed variables as evidence of criticality. Previous animal studies both in vitro and in vivo, together with computational modeling, have strongly suggested that the avalanche dynamics in neural systems may arise at the critical state in excitation–inhibition balanced networks and can be regulated by several intrinsic network properties, such as short-term synaptic plasticity and the balance level between excitation and inhibition~\cite{beggs2003neuronal, levina2007dynamical, lombardi2012balance}. With the help of simulating the stochastic dynamical model, we can fully understand the underlying mechanism of biological phenomena~\cite{sabir2022neuro, sabir2023fractional}. Jingwen's results demonstrate that the coordinated dynamics of criticality and asynchronous dynamics can be generated by the same neural system if excitatory and inhibitory synapses are tuned appropriately~\cite{li2020tuning}. In the excitation-inhibition balanced network, the critical state occurs at the point of Hopf bifurcation from a fixed point to a periodic motion~\cite{liang2020hopf}. In particular, neuronal avalanches can be triggered at an intermediate level of input heterogeneity, but heterogeneous output connectivity cannot evoke avalanche dynamics \cite{wu2019heterogeneity}. Notably, this finding is of particular interest because the coemergence of these multiscale cortical activities has been believed to ensure the cost-efficient information capacity of the brain, further emphasizing the functional significance of avalanche dynamics in neuronal information processing~\cite{yang2017co}.

\par
Recent studies have revealed that neurons communicating with each other require a fundamental understanding of neurotransmitter receptor structure and function~\cite{smart2012synaptic}. In a unified assumption that the cortical neuronal network is sustained in an E-I balanced state, network interactions can suffer much variability due to the difference in fast and slow-acting synapses. Theoretically, the variability in the synaptic connection pattern leads to stochasticity at the population level, which may further affect the spatiotemporal patterns of collective firing activity. Such stochastic effects indicate that different synaptic connections may be a potential factor in the regulation of neuronal avalanches. Another essential point is the background input, while in network modeling, it may be responsible for reproducing the avalanche criticality~\cite{renart2007mean, girardi2020synaptic}. If a cortical neuron works as an integrator over a relatively long time scale, it receives a substantial amount of excitatory drive, which can be referred to as a mean-dominated current. This means that the neuron is effectively summing inputs from a large number of sources over an extended period of time. On the other hand, when the mean drive is significantly smaller than the firing threshold, neurons can be activated by large fluctuations in the external current, leading to increased irregularity in their firing patterns. It remains controversial what context and network structure could contribute to evoking critical dynamics in the neuronal network.

\par 
Our paper aims to provide an intuitive, compact description of the spontaneous and critical state of cortical cells in terms of network coupling at mean-dominated and fluctuation-dominated afferent currents. To address this issue, we begin with a computing biological neuronal network in which spike communication contains both fast and slow synaptic types. In the mean-dominated regime, we find a clear transition of asynchronous firing to a synchronous state in the parameter space along with a power-law avalanche criticality. Different from the DP universality critical class, the model here reveals exponents with $\tau_{s}=1.6$ and $\tau_{t}=2.1$ through a maximum-likelihood estimator (MLE). We emphasize that the three network stages may be responsible for the avalanche criticality: mean-dominated subthreshold dynamics, synchronous initiation of a spike event, and delayed termination of the burst. Consistent with previous studies, we demonstrated that avalanches of neuronal activity preferentially emerge at a moderately synchronized state of collective firing activity, in which neurons are positioned around a hopf bifurcation point. With the presence of noise, the spiking activity passes through the bifurcation point and that network produces an occasional oscillation and a low stable firing, resulting in a moderate synchronous state. Finally, we implement a data assimilation method to fit the model to its BOLD signal and identify network parameters of criticality. These findings thus provide a new perspective to understand neuronal network criticality and enable researchers to model a biologically plausible network with the existence of criticality.

\section{Model and methods}
\label{sec: Methods}
\subsection{Neuronal network}
The network is composed of $N_{E}$ excitatory (80\%) and $N_{I}$ inhibitory (20\%) neurons~\cite{abeles1991corticonics}. Each neuron receives $K$ synaptic contacts from excitatory neurons and inhibitory neurons in a randomly connected manner, causing a 4/1 ratio of E/I synapses in the local network~\cite{knott2002formation}. Each node in the network is represented as a leakage integrate-and-fire (LIF) neuron and contacts through some biophysical synapses~\cite{kanagaraj2022effect}. 
\par 
This computational neuronal model is a nonlinear operator from a set of input synaptic spike trains to an output axon spike train, described by three components: the subthreshold equation of membrane potential that describes; the transformation from the synaptic currents of diverse synapses; the synaptic current equation that describes the transformation from the input spike trains to the corresponding synaptic currents; and the threshold scheme that gives setup of the condition for triggering a spike by membrane potential. This type of neuronal model has been demonstrated that tends to settle in a stationary fixed point, typically characterized by a stable pattern of firing activity~\cite{deco2012ongoing}. The dynamics of subthreshold membrane potential $ V(t) $ for excitatory (inhibitory) neuron $j$ can be described as

\begin{equation}\label{eq:cv_equation}
	\left\{
	\begin{aligned}
		&C_{j} \dot{V_{j}}  = -g_{L, j}(V_{j} - V_{L})- \sum_{u}S_{j,u}(V_{j} - V_{u})g_{i, u} + I_{ext}, \\
		&\dot{S_{j,u}} =-\frac{1}{\tau_{u}}S_{j, u} + \sum_{m, k}\omega_{m}\delta(t - t_{m}^{k}).
	\end{aligned}
	\right.
\end{equation}
\par 
Herein, the neuronal capacitance of neuron $j$ is denoted by $C_{j}$, and an external current $I_{ext}$ is applied as a background driver. When the membrane potential reaches the firing threshold $V_{th}$, the neuron emits a spike and its membrane potential is reset to $V_{reset}$ for a refractory period $T_{ref}$. $g_{L}$ and $g_{u}$ are the leak and synaptic conductance with equilibrium potentials $ V_{L} $ and $ V_{u} $ respectively. The gating variable $S_{j}$ represents the fractions of open channels of synaptic type $u$ with a decay time constant $\tau_{u}$. Postsynaptic currents are mediated by fast-acting (AMPA) and slow-acting (NMDA) excitatory and inhibitory GABAergic synaptic receptor types~\cite{brunel2001effects,deco2012ongoing}. These three synaptic types are abbreviated as $u=\{f, s, i\}$ for compact presentation. The $\delta$ function represents the incoming spike train and $t_{m}^{k}$ denotes the $k$th spike from the $m$th neuron. To account for network heterogeneity, we sample the synaptic weight $\omega_m$ from a uniform distribution. 

\par 
To simulate external input in addition to local neuronal network activity, we incorporate an Ornstein-Uhlenbeck (OU)-type current with varying parameters ($\mu_{ext}, \sigma_{ext}, \tau_{ext}$) into our model, 
\begin{equation}
    \label{eq:ou}
    \tau_{ext}~d I_{ext}=(\mu_{ext} - I_{ext})dt + \sqrt{2 \tau_{ext}}\sigma_{ext}dW_{t}.
\end{equation}
We will present a systematic discussion of spiking activities in a conductance-based neuronal network under two distinct current-driven scenarios. The default neuronal parameters and network settings used in this paper are shown in the following Table~\ref{tab: table1}.
\begin{table}[htbp]
	\centering
	\caption{\bf Default parameters and settings used in numerical simulation}
	\begin{tabular}{lll}
	\toprule
	Symbol & Description & Value\\
	\midrule
	$N$ & Total number of neurons & $2000$ \\
	$K$ &  Average number of input connection & $100$ \\
	$C$ & Capitance & $0.5$ nF\\
 	$T_{ref}$ & Refractory period & $2$ ms \\
 	$V_{th}$ & Voltage threshold & $ -50 $ mV\\
	$V_{reset}$ & Reset potential & $-55$ mV\\
	$g_{L}$ & Leaky conductance & $25$ nS\\
	$\tau_{u}$ & Decay time of synaptic receptor & $(2, 40, 10)$ ms \\
	$V_{u}$ & Reverse potential of different receptor & $(0,0 -70)$ mV\\
	$V_{L}$ & Equilibrium potential of leak & $ -70$ mV\\
        $\omega_{m}$ & Synaptic weight  & Uniform[0, 1] \\
        $\mu_{ext}$ & Mean level of OU current & $0.45 \,\mu$A \\
        $\sigma_{ext}$ & Fluctuation of OU current & $0.15 \,\mu$A\\
        $\tau_{ext}$ & Correlation time of OU current & $2.5$ \\
	\bottomrule
	\end{tabular}
    \label{tab: table1}
\end{table}

\subsection{Balloon model}
The blood-oxygen-level-dependent (BOLD) signal reflects changes in deoxyhemoglobin driven by localized changes in brain blood flow and blood oxygenations, which are coupled to underlying neuronal activity by a process termed neurovascular coupling. In this paper, we adapt the Balloon-Windkessel model~\cite{friston2000nonlinear} for BOLD signals and model them as observation in the data assimilation~(\nameref{sec: da}). The Balloon model is based on a set of ordinary differential equations that describe the hemodynamic response to changes in neuronal activity. Neuronal activity $z$ drives the responses of cerebral blood flow $s$ and rate of deoxyhemoglobin $q$ and outputs the BOLD signal $h$. The model architecture is summarized as follows:
\begin{equation}
\label{eq: balloon}
\left\{\begin{array}{l}
\dot s=\varepsilon z-\kappa s-\gamma\left(f-1\right) \\
\dot{f}=s \\
\tau \dot{v}_i=f-v^{\frac{1}{\alpha}} \\
\tau \dot{q}=\frac{f\left(1-(1-\rho) \frac{1}{f}\right)}{\rho}-\frac{v^{\frac{1}{\alpha}} q}{v} \\
h=V_0\left[k_1\left(1-q\right)+k_2\left(1-\frac{q}{v}\right)+k_3\left(1-v\right)\right]
\end{array}\right.,
\end{equation}
where $\epsilon=200, \kappa=1.25, \tau=1, \gamma=2.5, \alpha=5, \rho=0.8, V_{0}=0.02, k_1=5.6, k_2=2, k_3=1.4$ are constant parameters of the Balloon model and are chosen as in \cite{friston2000nonlinear, lu2022human}.

\par 
Furthermore, to mimic the time series of BOLD signals that were measured by the fMRI, we conducted a downsampling process of the recording data. The valid observation is sampled with a periodical duration of $800$ ms, which is the same as the frequency in the canonical experiment paradigm.

\subsection{Data analysis}
Simulations are done using a finite difference integration scheme based on the first-order Euler algorithm with a time step of $dt = 10^{-1}$ ms. Each network is simulated for a long time of $ 200$ s with the initial period of $20$ s discarded to record sufficient data. It is noteworthy that the specific time (sufficiently long) of data recording does not exert any influence on the results. To obtain convincing statistical measurements, we carry out 20 trials for each simulation configuration. Networks are simulated on a GPU card running Linux using custom-written codes in C++ and Python.

\paragraph{Irregularity of individual spikes}
The coefficient of variation (CV) is a commonly used measure in the analysis of spike data from neuronal networks. The CV is a statistical measure of the variability of the interspike intervals (ISIs) and is defined as
\begin{equation}
	\mathrm{CV}_{\mathrm{ISI}}=\left\langle\frac{\sqrt{\left\langle T_{j}^{2}\right\rangle-\left\langle T_{j}\right\rangle^{2}}}{\left\langle T_{j}\right\rangle}\right\rangle .
\end{equation}
Here the symbol $\langle\cdot\rangle$ represents the average and $T_{j}$ is the inter-spike interval of neuron $j$. A low CV indicates a regular firing pattern, where the ISIs are relatively constant, while a high CV indicates an irregular firing pattern, where the ISIs are more variable. 

\paragraph{Synchrony index}
The coherence coefficient (cc) is a commonly used measure to assess the synchronization of population activity in neuronal networks~\cite{wang2002pacemaker, wu2019heterogeneity}. It quantifies the degree of phase locking between a pool of neurons by accessing the macroscopic firing rate. In our model, the instantaneous population firing rate $ z (t) $ is calculated in each $0.1$ ms bin. Then, we can define the dimensionless measurement as the ratio of the standard deviation and mean:
\begin{equation}
	\mathrm{CC}=\frac{\sigma_{z(t)}}{\mu_{z(t)}}.
\end{equation}
Obviously, the larger the value of cc is, the better the synchronization in the network.

\paragraph{Oscillation power}
In our investigation of the network, we are studying the oscillation power by analyzing the power spectrum of the time series of the population firing rate. To obtain the power spectrum, we used the Welch method with a Hamming window of size $2^{11}$ and performed a fast Fourier transform (FFT) with the same number of points. By scrutinizing the curve of power spectral density, we can ascertain both the acme power amplitude of neuronal oscillations and its corresponding peak frequency.

\paragraph{Avalanche dynamics}
In agreement with previous studies on spike-based neuronal avalanches in vivo~\cite{bellay2015irregular}, we defined the neuronal avalanches only using spikes in the excitatory population. First, we sample the spike data from $ N_{S} $ randomly selected excitatory neurons and bin the population activity into time windows ($ \Delta t = 0.5 ms $). Then, the threshold $\Theta$ to determine the start and end of an avalanche event is defined as the 50\% median spiking activity of the excitatory population. A neuronal avalanche starts and ends when the summed activity exceeds the $\Theta$. For a given avalanche event, we denote its size $ S $ and duration $ T $ as the total number of spikes contained in this event and the corresponding lifetime of this event. By using all avalanche events recorded in a specific experiment, we can estimate the probability distributions of avalanche size and duration, represented by $ P (S) $ and $ P (T) $, respectively.
\par 
To characterize neuronal avalanches, the distribution $ P(s) $ of avalanche sizes is first visually inspected and then quantified by the $\kappa$ index~\cite{shew2009neuronal}, which calculates the difference from an observed (simulated) distribution $P^{obs}$ with the known theoretical power-law distribution $ P^{th}(s) $, at 10 equally spaced points on a logarithmic axis and adds 1:
\begin{equation}
	 \kappa=\frac{1}{10}\sum_{i=1}^{10} \left(P^{obs}(s_{i})-P^{th}(s_{i})\right) + 1.
\end{equation}
Within this definition, a subcritical distribution is characterized by $\kappa < 1$, and a supercritical distribution by $\kappa > 1$, whereas $\kappa = 1$ indicates a critical network. The theoretical avalanche criticality discussed here is the true MF-DP critical point with scale-invariant avalanche distributions that obey the cracking noise scaling relation~\cite{girardi2020synaptic}. As mentioned previously, $\tau_{s} =3/2$ and $\tau_{t} =2$ stand for MF-DP models at or above the upper critical dimension.

\subsection{Data assimilation}
\label{sec: da}
An augmented Ensemble Kalman filter (EnKF) is applied here to identify the dynamical connection by incorporating the synaptic parameters as an augmented state vector~\cite{lourens2012augmented}. For assimilation purposes, The joint input-state estimation can be achieved by including the unknown forces in the state vector and estimating this augmented vector using a standard Kalman filter. We rewrite the state-space model with an $1+5+1$ dimensional state vector $\Upsilon_{t}$ and a $1$-dimensional observation vector $h_{t}$ according to
\begin{equation}
\begin{aligned}
&\Upsilon_{t} = \mathcal{F}(\Upsilon_{t-1}) + \sigma_t, \quad \sigma_t \sim \mathcal{N}(0, Q), \\
&h_t = \mathcal{G}(\Upsilon_t)+\epsilon_t , \quad \epsilon_t \sim \mathcal{N}(0, R),
\end{aligned}
\end{equation}
where $w_t$ and $v_t$ are Gaussian noise terms with covariance matrices Q and R respectively. Herein, the function $\mathcal{F}$ is composed of the neuronal network, balloon model and random walk model of the estimated parameter,
\begin{equation}
    \Upsilon_{t} = \left(
    \begin{array}{c}
         z_{t}  \\ y_{t} \\ \theta_{t}
    \end{array}
    \right),
    \quad
    \mathcal{F}(\Upsilon_{t-1})=
    \left(
    \begin{array}{c}
         \textit{NN}(\theta_{t-1}) \\ \textit{Balloon}(y_t, z_t) \\ \theta_{t-1} + \eta\xi_{t-1}
    \end{array}
    \right),
\end{equation}
where 1-dimensional $\theta_{t}$ and 1-dimensional $z_{t}$ denote the parameter and the population firing rate in the evolution process. The $5$-dimensional $y_t$ represents the hidden states that emerged in the Balloon model in equation~\ref{eq: balloon}, which together contribute to the BOLD signal. The state vector is directly related to the force parameter $\theta$ with a stochastic process. In agreement with previous studies, the observation $\mathcal{G}$ is simply designated to be a linear map, which projects the state vector to its one element $h_{t}$.
\begin{equation}
    \mathcal{G}(\Upsilon_{t}) = \left(\begin{array}{ccccc}
         0 & \ldots & 1 & \ldots & 0 \\
    \end{array}
    \right) \cdot  \left(
    \begin{array}{c}
         z_{t}  \\ y_{t} \\ \theta_{t}
    \end{array}
    \right).
\end{equation}
\par 
In the DA framework, we adopt a reversible map to map $\theta_{t}\in(a, b)$ to an infinite range $\Theta_{t} \in (-\infty, \infty)$ to avoid exceeding the bound of the parameter range while applying a random walk. We set the map function $\Phi$ as a $log$ function and its corresponding reversible map is $sigmoid$. Under this condition, the controlling parameter $\theta$ is fed into the original model after the prediction and updating procedure. The complete algorithm is as follows (Algorithm~\ref{EnKF}):
\renewcommand{\thealgorithm}{}
\begin{algorithm}\label{EnKF}
	\caption{EnKF}
	$\mathbf{Input:}$\\
	BOLD signals $h_{t},t=1\ldots T$;\\
	number of total samples $N$; total time length $T$;\\
	init $N$ networks with different parameters $\theta$;\\
	observation noise $R$; random walk step $\eta$\\
	$\mathbf{Output:}$\\
	estimated BOLD signals;\\
	hidden states and parameters;\\
	\label{ABCLFRS}
	\begin{algorithmic}[1]
		\State Draw $N$ samples $\theta_{0}^{n}$ from initial distribution;
        \State map $\theta_{t}$ to $\Theta_{t}$ through function $\Phi$.
		\For{$t=1:T$}
		\State apply random walk as $\Theta_{t}=\Theta_{t-1}+\eta \xi_{t-1}$;
        \State update the parameter $\theta$ through the function $\Phi^{-1}$ to the evolution model.
        \State update the hidden state includes the neuronal states and hemodynamical states.
		\State evolve each neuronal network and obtain the bold state $\hat{h_{t}^{n}}$, then concatenate to $\hat{\Upsilon_{t}^{n}}$.
		\State calculate $\mu_{t}=\frac{1}{N}\sum \hat{\Upsilon_{t}^{n}}$, $C_{t}=\frac{1}{N-1}\sum(\hat{\Upsilon_{t}^{n}}-\mu_{t})(\hat{\Upsilon_{t}^{n}}-\mu_{t})^{T}$.
		\State derive Kalman gain matrix $S_{t}=HC_{t}H^{T}+R$, $K_{t}=C_{t}H^{T}S_{t}^{-1}$
		\If{$t\%800==0$}
			 \State Draw $\epsilon_{o}^{n}$ from $\mathcal{N}(0, R)$, filter by $\Upsilon_{t}^{n}=\hat{\Upsilon_{t}^{n}} + K_{t}(y_{t}+\epsilon_{o}^{n}-H\hat{\Upsilon_{t}^{n}})$
		\EndIf
		\EndFor
	\end{algorithmic} 
\end{algorithm}

\section{Results}
\label{sec: Results}
\subsection{Transcritical state}
To investigate the behavior of the system, we conduct a preliminary study using a small network consisting of 2000 neurons, under two scenarios characterized by different levels of driven current. One is mean-dominated and the other is fluctuation-dominated type, which is reflected by variations in the parameter pair $\mu_{ext}$ and $\sigma_{ext}$. We ask whether the fast-acting coupling strength contributes to the modulation of network dynamics. In our model, the 2 free variables, $\rm AMPA$ and $\rm NMDA$ type synaptic conductance, are constrained by a linear equation to better explore spiking patterns in a locking firing regime~(\nameref{sec: mechanism}). We calculate various neuronal measurements to gain insight into the properties of the network.
\begin{figure}
\centering	
\includegraphics[width=0.9\linewidth]{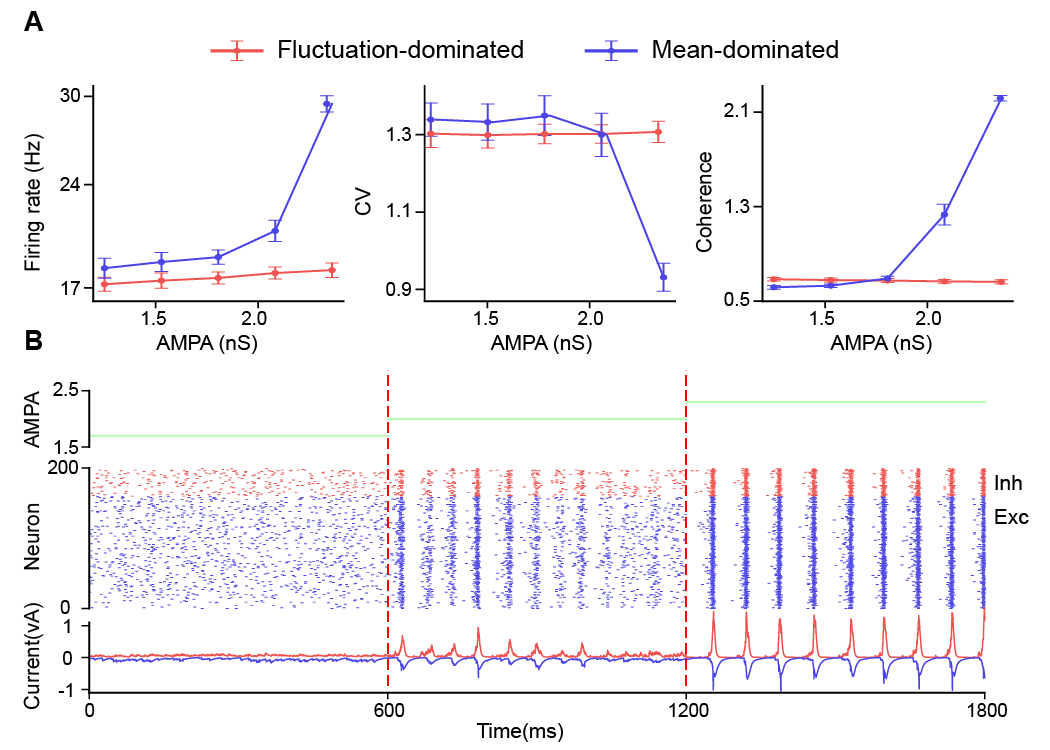}
\caption{{\bf Spiking activity with respect to the excitatory conductance}
(A):  Network activity (firing rate, coefficient of variance and coherence coefficient with respect to different $\rm AMPA$ conductance while keeping the constraint $\rm AMPA + 20 \times NMDA = 2.5$. There exists an obvious sharp transition from an asynchronous to a synchronous state in the one-dimensional transition, in which the boundary is characterized as a moderately synchronized state, as shown in  the middle panel of (B).
(B): The local recurrent neuronal network consists of Exc. and Inh. spiking neurons with different $AMPA$ levels. Network activity of three typical points with parameters indicated in the top panel. middle, raster plot of a subset of 200 neurons (Exc. 160 (black), Inh. 40 (red)); bottom, the average excitatory and inhibitory synaptic currents.}
\label{fig1}
\end{figure}
\par
We begin with the fluctuation-dominated regime in which the external drive, as described by the mean of the OU current, $\mu_{ext}$, and the deviation of each step, $\sigma_{ext}$, has a high mean and a small deviation. With the linear constraint of excitatory synaptic conductance, the firing rate of the network is located at approximately 17 Hz. In the fluctuation-dominated case, all CV values are much larger than 1, indicating an irregular firing pattern. Although strengthening the AMPA conductance changes the network structure, it results in almost no additional variability in neuronal firing, which is quantitatively confirmed by CV and coherence coefficient (Fig~\ref{fig1} A).
\par 
In the mean-dominated regime, a strong external drive provides the mechanism for a majority of the subthreshold voltages to remain close to one another. In this way, the mean drive to the cell was subthreshold, and spikes were the result of fluctuations, which occur irregularly, thus leading to a high CV. Increasing the AMPA conductance in the network significantly enhances the regularity of spike activity (Fig~\ref{fig1} A). Such a strong fast-acting synapse could integrate pre-spikes rapidly and trigger a subset of neurons firing rapidly at the same time. A cascading firing event occurs when all neuron voltages are positioned sufficiently close to the firing threshold such that some spikes in the network are sufficient to induce firing in all other neurons in the network at that same time. Such synchronized behavior might also enhance the effect of collective firing activity, and therefore, an increase in the average firing rate appears when there is a high level of AMPA conductance.

\par 
From Fig~\ref{fig1} B, we can see that both synchronous and asynchronous irregular firing patterns are observed in this same network. Moderate synchrony with a moderate correlation (0.8-1.8) appears in the intermediate AMPA efficacy. In the moderately synchronized case, neuronal avalanches are organized rhythmically into bursts, large and small, separated by distinct silences. The excitatory current is induced quickly, followed by a long period of inhibitory current. Neurons show sparsely synchronized oscillations, which consist of irregular and sparse individual spikes but synchronized oscillating population activities. 

\subsection{Rich spiking patterns}
\begin{figure}
\centering	
\includegraphics[width=0.9\linewidth]{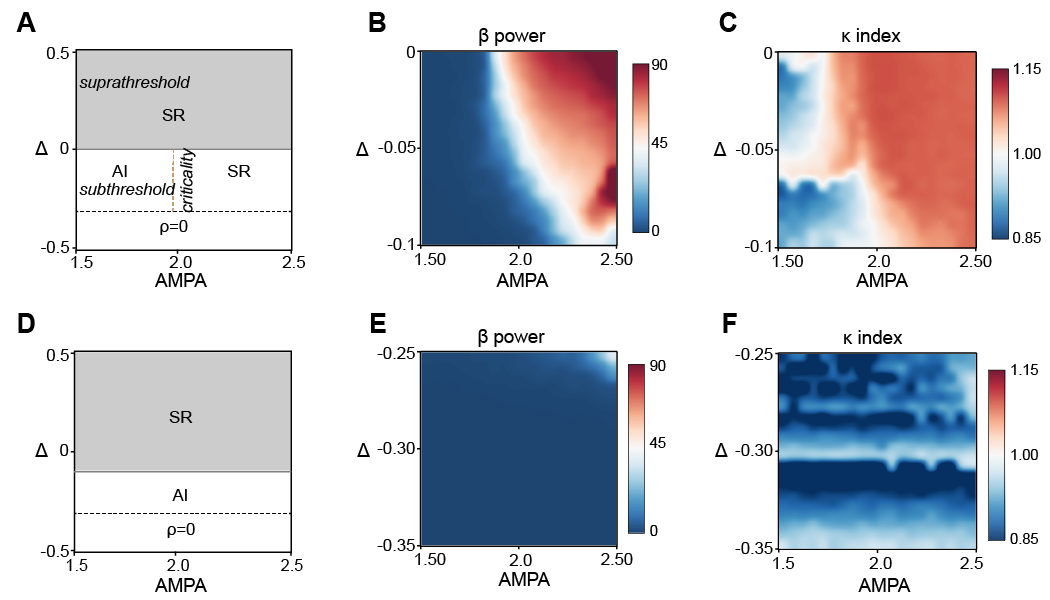}
\caption{{\bf phase diagram and spiking avalanches of the neuronal network.} (A, B, C): The phase diagram of the network with respect to $\Delta$ and $\rm AMPA$ conductance in the mean-dominated driven case. The parameter space suggests a phase transition, as shown in the heatmap of (C) the $\kappa$ coefficient and (E) the power of collective oscillations. (D, E, F): the same illustration as the top panel but in the condition of fluctuation-dominated current.}
\label{fig2}
\end{figure}

In the LIF model, the input of the synaptic current depends on two aspects: first is the external current and the interactions in the local network. In the noise-free network, neurons can fire sustainably only when the external stimulus is of the order of the threshold potential, $I^{ext}\sim V_{th}$. As in~\cite{girardi2021unified}, we define the parameter $\Delta=I^{ext} - V_{st}$ as the external suprathreshold current (details in~\nameref{sec: Methods}). The summary of phase transition undergone in two different types of OU current is given in Fig ~\ref{fig2} A and D.

In the suprathreshold regime $\Delta > 0$, the limited stable potential $V_{st}$ is larger than the threshold potential, leading to synchronous events and regular spiking activities. Note that the suprathreshold spiking activities can also exist in the narrow region of $\Delta <0$ in the fluctuation-dominated regime. In this region, $V_{st}$ is close to the firing threshold, and a sufficiently large fluctuation of input current could induce firing in most neurons in the network. Since all neurons spike at the same time, their voltages are all reset to rest at the same time, causing periodic synchronous bursts and a high firing rate. If $\Delta << 0$, neurons are silent with zero average firing rate $\rho =0$. Interestingly, in the subthreshold regime $\Delta ~ 0$, the neurons can undergo a transition from an asynchronous irregular (AI) to a synchronized regular (SR) state with an increase of impact in the fast-acting synapses under a mean-dominated input. Excitatory and inhibitory neurons in the network receive mean-dominated input that drives their voltage close to the threshold and gives rise to synchronized cascading firing events. The model described in equation~\ref{eq:cv_equation} assumes that the fast-acting synaptic conductance is much larger than the slow one, allowing for rapid integration of spikes from pre-synapses and the generation of strong, pulse-like currents. A cascading firing event occurs when a large and rapid excitatory current induces a chain reaction of spike firing, resulting in all other neurons firing at the same time. Then, the inhibitory neurons become active to suppress the firing event for a short period, which leads to the emergence of network oscillations. At the boundary line of AI and Si, the network exhibits totally different spiking activities: synchronous cascading events do not occur periodically, but rather at random times, accompanied by large or small silent windows. This moderate synchrony among neurons may emerge when the network is operating near a critical point~\cite{chialvo2010emergent}. Nevertheless, a similar transition behavior does not appear in networks with fluctuation-dominated external input. These findings suggest that mean-dominated input is essential in the emergence of neurodynamics.
\par
\begin{figure}
\centering	
\includegraphics[width=0.8\linewidth]{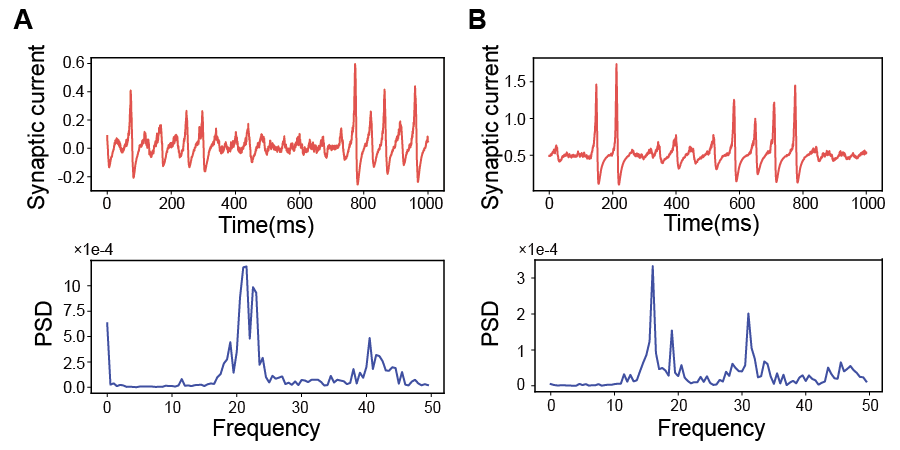}
\caption{{\bf Inhibition modulates the oscillation frequency.} (A) Oscillations in the model of criticality and PSD of the synaptic current for the inhibitory time constant $\tau_{i}=10$ ms. (B) The different oscillations with much smaller frequencies are shown on the right holding $\tau_{i}=12$ ms for all neurons. }
\label{fig3}
\end{figure}

 
Then, in the subthreshold regime, we assessed the beta power of the firing rate in two different situations. From Fig~\ref{fig2} E, the synaptic interaction does not regulate neuronal behavior in the fluctuation-dominated regime as mentioned above. However, in the mean-dominated regime, depending on the fast-acting synaptic conductance, the networks exhibit oscillations in the $\beta$ band (13-30 Hz range) along with a clear transcritical line. Synchronous excitatory neurons fire an action potential, exciting the inhibitory neurons to fire, in turn suppressing the excitatory neurons from firing for a period of time. This interaction between excitatory and inhibitory spiking is the fundamental principle underlying the pyramidal-interneuronal network gamma (PING) type oscillation~\cite{borgers2005effects}. The inhibition synapse plays an important role in regulating the frequency of the oscillations by noting that the power spectral density (PSD) shifts to the left with the increase in the decay time of the inhibition channel (fig~\ref{fig3}). We verify that the oscillations stem by plotting the typical spiking activity of the inhibition and excitation. Inhibition plays an essential role in modulating the frequency of the oscillations by cancelling the excitatory current with an induced inhibitory current.

\par
Furthermore, we investigate the $\kappa$ index in the subthreshold regime, to obtain an assessment of how close the observed avalanche distribution close to the Direct Percolation (DP) universality class~\cite{munoz1999avalanche} for exponents $\tau=3/2$ and $\tau_{t}=2 $ respectively. We found that the $\kappa$ index is smaller than 1 on the left side of the subcritical regime, corresponding to the subcritical regime. In contrast, the network exhibits supercritical phenomena when fast-acting excitation is large enough with $\kappa > 1$. In the transcritical line, the network obeys the $\kappa = 1$ corresponding to the DP critical network. Interestingly, the region of $\kappa = 1$ (Fig~\ref{fig2} C) does not correspond well with the boundary line of the $\gamma$ power distribution (Fig~\ref{fig2} B), in which more patches of  $\kappa =1 $ emerge in the diagram. The above-presented results suggest that DP universality criticality is not the most appropriate class in the transition to collective oscillations. Henceforth, we relax this constraint, taking a more agnostic approach towards the values of the exponents and letting the MLE method determine them (~\nameref{sec: avalanches}). The transcritical line is not vertical in the subcritical regime, which implies that the critical parameter of $g_{\rm NMDA}$ is not fixed in the different external current levels.

\subsection{Neuronal avalanches}
\label{sec: avalanches}
\begin{figure}
\centering
\includegraphics[width=0.9\linewidth]{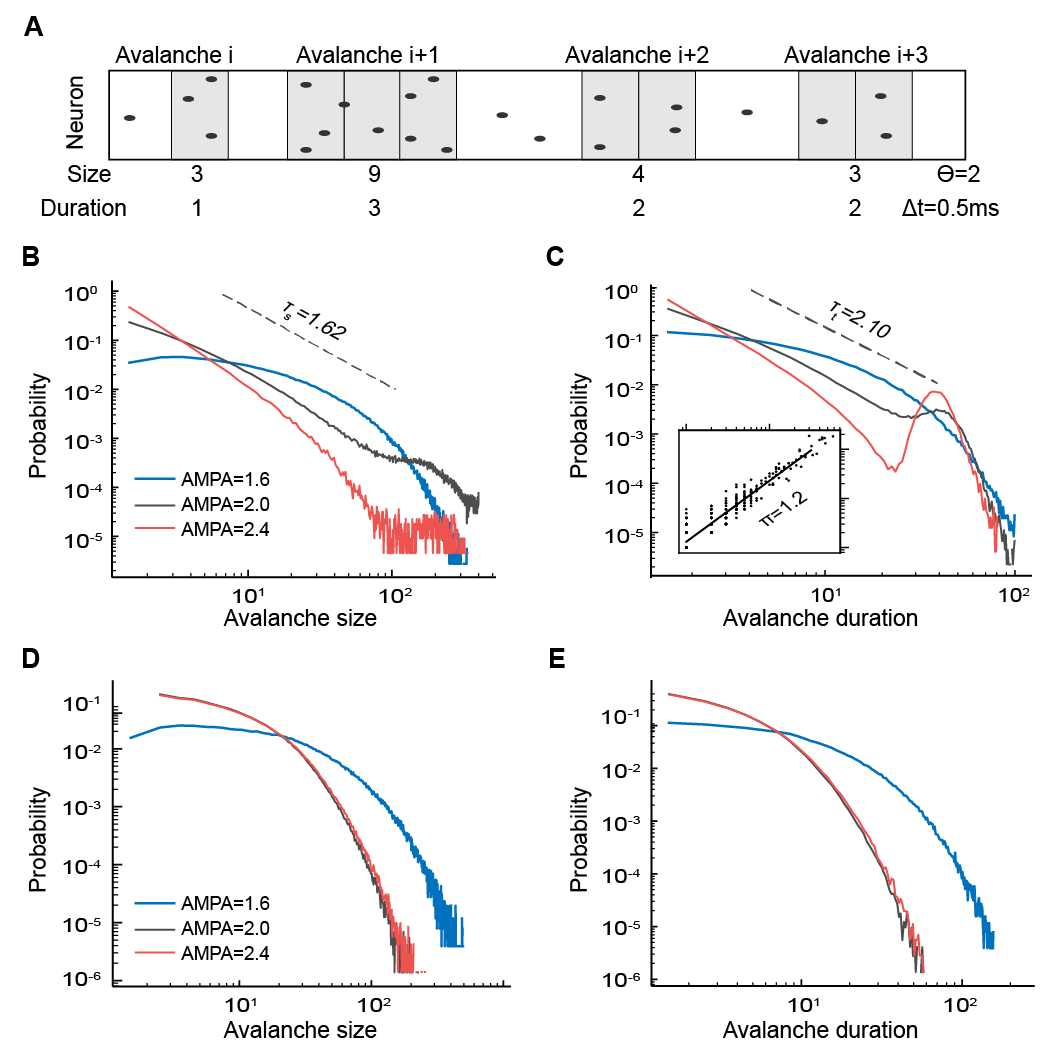}
\caption{{\bf Critical state with spike avalanches.}
(A): mapping spikes from $N_{s}$ randomly sampled excitatory neurons into time bins ($\Delta t = 0.5$ ms). Here an avalanche event is defined as a sequence of time bins in which the spiking count at least exceeds $\Theta$, ending with a "silent" time bin.
(B, C): Typical distributions of avalanche size and avalanche duration for networks with different synaptic parameters in different regions. At different levels of $\rm AMPA$ conductance, the model may present subcritical (red), critical (black), and supercritical (blue) avalanche dynamics. Inset: at the critical state, the average size $S$ conditioned on a given duration $T$ shows power-law increases corresponding to $S \sim T^{\pi}$. (D, E): The corresponding distributions of avalanche events for networks in fluctuation-dominated regime. Similarly, three levels of synaptic conductance are considered and the model only exhibits subcritical dynamics due to high fluctuation among neurons.}
\label{fig4}
\end{figure}

To further re-examine the critical phenomena, we implement a simulation with a much longer time $T_{sim} = 500 s$ and calculate the spike-based size and duration of each avalanche event (\ref{fig4} A). For comparison with real electrophysiological experiments, we collect avalanche events from a small population of randomly chosen excitatory neurons (the sampling size $N_{s}=400$ for default), but not all neurons, in the network.
Fig~\ref{fig4}A and B show the distribution of avalanche size and avalanche duration for networks in the three typical examples in Fig~\ref{fig1}. The model exhibits distinct avalanche dynamics at different levels of fast-acting coupling strength. The subcritical dynamics have an exponentially decaying avalanche distribution(~\ref{fig4} blue lines); the supercritical dynamics have a much greater chance of large-size avalanches(~\ref{fig4} red lines), while the critical dynamics show a power-law avalanche distribution(~\ref{fig4} black lines). In the critical state, the distributions of both avalanche size and duration obey linear relationships in log–log coordinates. These two linear relationships can be well characterized by the exponents of power-law statistics ($\tau_{s}=-1.84$ and $\tau_{t} = -2.10$), which are reasonably close to the value for the DP point. Remarkably, the further plotting of the average avalanche size as a function of avalanche duration in the log–log coordinates reveals another power-law scaling, with an exponent of $\pi$ = 1.2 (Fig~\ref{fig2}C inset). These exponents satisfy the expected relation $\pi = (1 + \beta) /(1 + \alpha)$ as predicted for a critical system by the scaling theory of nonequilibrium critical phenomena~\cite{rybarsch2014avalanches}. In agreement with previous in vivo and in vitro experiments, these findings together suggest the occurrence of avalanche dynamics in our model.
\par 
However, a similar tuning effect is not observed in networks with fluctuation-dominated external input(Fig~\ref{fig4} D, E). The high fluctuation of input may lead to asynchronous irregular neuronal firing, triggering subcritical dynamics with exponential-type distributions of avalanche size and duration. Theoretically, this finding is not surprising because variability in fast-acting synapses has been shown to contribute weakly to collective neuronal firing in the fluctuation-dominated regime.
\par
\begin{figure}
\centering	
\includegraphics[width=0.8\linewidth]{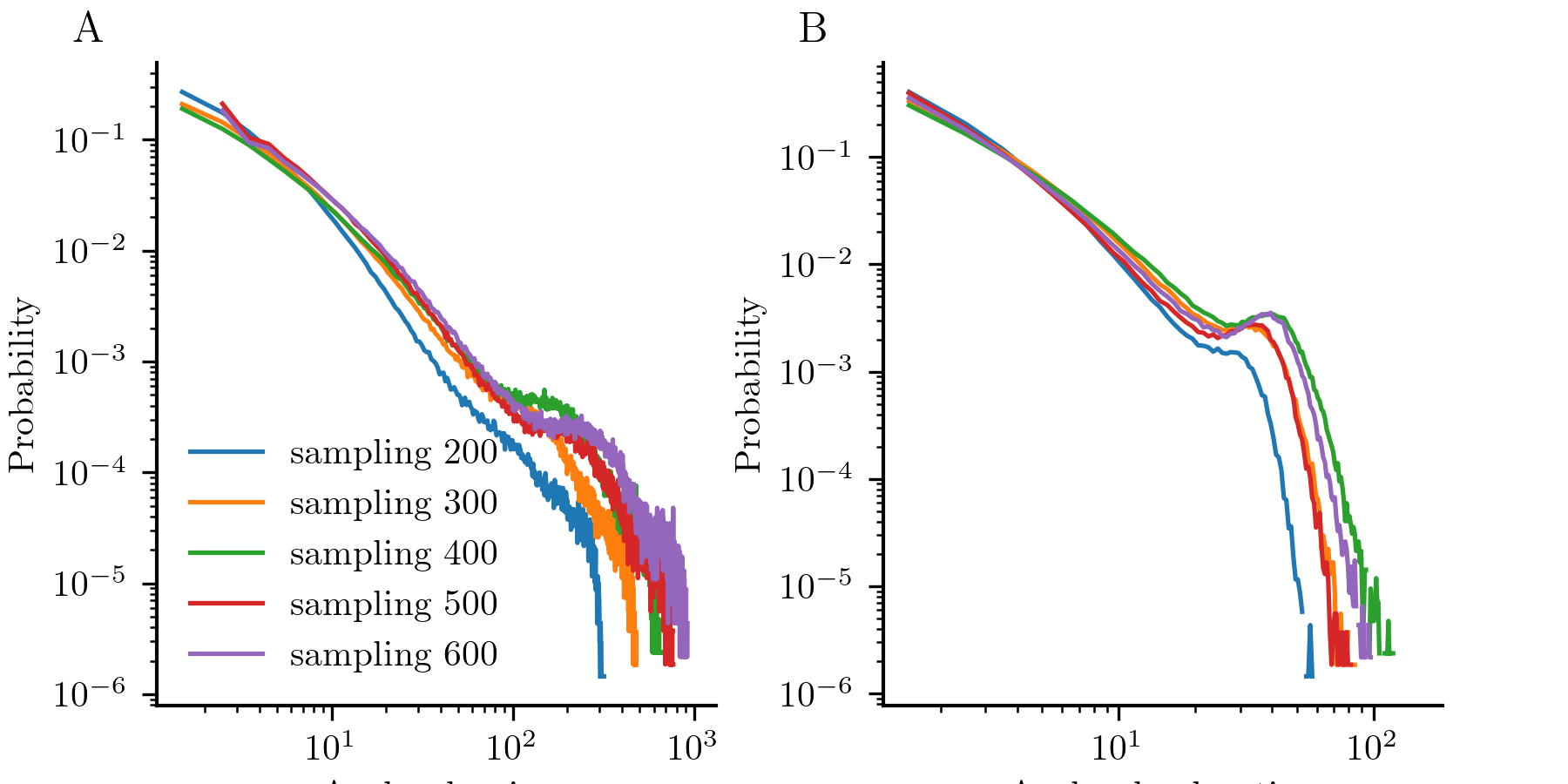}
\caption{{\bf Critical exponents vary in different recording sizes.} (A): Avalanche assessment in different sampling sizes. (B) Corresponding distributions of avalanche duration at different sampling sizes.}
\label{fig5}
\end{figure}

 
Furthermore, we investigate the question that whether the sampling size of neuronal data influences the power-law distribution of neuronal avalanches (Fig~\ref{fig5}). Using the same spike data generated in the mentioned critical setting, we recalculate the distributions for both avalanche size and duration at different sampling sizes. We find that the power-law statics of avalanche events can be preserved, illustrating the stability of the neuronal avalanches generated in our model. Recording insufficient spike-based events from a limited number of neurons does not prevent the observation of the characteristic power law. However, large population recording remarkably increases the possibility of large-size avalanches, which may be responsible for the much smoother slope in avalanche distribution. This observation implies that the exponents of the power-law avalanches may vary in different sampling experimental paradigms.

\subsection{Mechanism of criticality}
\label{sec: mechanism}
To gain further mathematical insight into the existence and modulation of the criticality, we coarse-grain the LIF model to a firing-rate model for this network, writing equations for the three gating variables: fast excitatory $S_{f}(t)$, slow excitatory $S_{s}(t)$ and the inhibitory variable $S_{i}(t)$. Then, the time-dependent neuronal firing rates $\rho(t)$ for the network are derived in terms of these network gating variables, which are then fed back into the evolution of the network conductance variables, closing the system.
\par 
As a preliminary approximation, we assume that $V_{j}(t)$ obeys a Gaussian distribution $p(V, t)$ with a time dependent mean $V_{0}(t)$ and time-independent variance $\sigma_{0}(t)$. In agreement with the previous analysis framework, we suppose that the neuronal potential $V_{i}$ is restricted to $(-\infty, V_{th})$ and that $V_{th}$ is an absorbing boundary. Considering the $N$ neuron voltages as independent between two spike events, the neuronal firing rate of the network can be formulated as the expectation of the proportion of the neurons whose membrane potential is above the spiking threshold~\cite{liang2020hopf}. Then, 
\begin{equation}
\label{eq: estimation}
\begin{aligned}
\rho(t)=\left\langle H\left(V(t)-V_{t h}\right)\right\rangle & =\int_{V_{t h}}^{\infty} p(V, t) d V \\
& =\frac{1}{2}-\frac{1}{2} \operatorname{erf}\left(\frac{V_{t h}-V_{0}(t)}{\sqrt{2} \sigma_{0}(t)}\right),
\end{aligned}
\end{equation}
\par
Suppose the pulses that arrive from the excitatory population and inhibitory population are assumed to be a Gaussian distribution with a high firing rate $\rho(t)$. Thus, the drive term $S_{u}$ can be described as a constant term and Gaussian white noise, and is governed by
\begin{equation} 
\label{eq: gating_ou}
    \tau_{u} \frac{d S_{u}}{d t} = \omega \tau_{u}\rho -S_{u}+ \omega \tau_{u} \sqrt{\rho}\xi(t),
\end{equation}
\par 
where the Gaussian white noise $\xi(t)$ has a mean and auto-correlation function defined by
\begin{equation}
    \langle \xi(t)\rangle = 0 ,\quad \langle \xi(t) \xi(t^{\prime})\rangle=\delta(t-t^{\prime}).
\end{equation}
This OU process has been shown to capture the statistics of conductance fluctuations at some of the compartmentalized model neurons\cite{destexhe2001fluctuating}. In fact, the criterion for the validity of the diffusion approximation is that the mean value is much larger than the fluctuation term. By noting that the time constant of slow excitation and inhibition is much larger than that of fast excitation, we can neglect the fluctuations of these two gating variables and approximate them as 
constant coefficient differential equations. 
\par 
Using the effective-time constant diffusion approximation~\cite{richardson2005synaptic}, the equation~\ref{eq:cv_equation} reduces to
\begin{equation}
    \label{eq: approximation}
     \tau_{0}\frac{dV}{dt} = - (V - V_{0}) + \sigma_{0}\sqrt{\tau_{0}}\eta(t),
\end{equation}
where
\begin{equation}
    \label{eq: approximation_detail}
    \left\{
    \begin{aligned}
        & g_{0} = g_{L} + \sum_{u}g_{u}S_{u} \\
        &\tau_{0} = C / g_{0} \\
        &V_{0} = \frac{1}{g_{0}} * (g_{L}V_{L} + \sum_{u}g_{u}S_{u}V_{u}) + \mu_{ext} \\
        &\sigma_{0}=\frac{\sigma_{ext}}{g_{0}}\sqrt{\frac{2\tau_{ext}}{\tau_{0}}}.
    \end{aligned}
    \right.
\end{equation}
In equation~\ref{eq: approximation_detail}, the increased effective leak $g_{0}$ incorporates the effect of synaptic conductance, and $\tau_{0}$ denotes the effective time constant. The distribution predicted for the voltage is Gaussian, and the average and variance of the membrane potential are $V_{0}$ and $\sigma_{0} ^{2}$, respectively.
\par 
The second-order OU process governs the trace of the membrane potential in an ideal case, which is dependent on $\tau_0, V_0$ and $\sigma_0$ in equation~\ref{eq: approximation_detail}. In this paper, to gain deep insight into spiking patterns regardless of network firing rate, we approximately lock the firing frequency by constraining the two excitation conductances (AMPA and NMDA) to a linear equation as
\begin{equation}
    \label{eq: constrain}
    \langle g_{f}S_{f} + g_{s}S_{s}\rangle = g_{f}\omega \tau_{f}\rho + g_{s}\omega \tau_{s}\rho = \textit{const}.
\end{equation}
An appropriate coupling strength of the constant in equation~\ref{eq: constrain} provides a sufficient range of fast excitation in the experiment while keeping the firing rate approximated fixed (Fig~\ref{fig1}).
\begin{figure}
\centering
\includegraphics[width=0.8\linewidth]{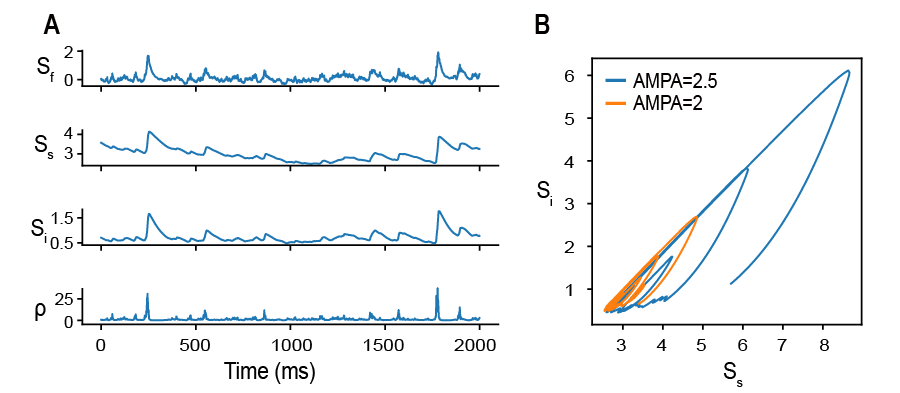}
\caption{{\bf Dynamics of the field model}
(A): The evolution of their gating variable and network firing rate in the parameter of power-law criticality. (B): The trajectories of $S_{s}$ and $S_{i}$ in the supercritical parameter and represented as a limit cycle.}
\label{fig6}
\end{figure}
\par 
We find that the involvement of the synaptic shot noise of fast excitation (equation~\ref{eq: gating_ou}) in the diffusion approximation does not influence the performance of the approximation (equation~\ref{eq: approximation}). The network is still stochastic due to the existence of a fluctuation term in the external current. Generally, mean-field theory only holds when the system size is infinity. The incorporation of noise into the field model can smooth out the systematic errors, compensate for the finite size effect and make it closer to the true rate dynamics statistically. Thus, for numerical simulation of the field equations we will keep the noise terms in equation~\ref{eq: gating_ou}. Finally, we close the system and arrive at the field equations as
\begin{equation}
\label{eq: closed_system}
    \left\{
    \begin{aligned}
    & \rho_{t} = \frac{1}{2}-\frac{1}{2} \operatorname{erf}\left(\frac{V_{t h}-V_{0}(t)}{\sqrt{2} \sigma_{0}(t)}\right) \\
    & \tau_{f} \frac{d S_{f}}{d t} = \omega \tau_{f}\rho -S_{f}+ \omega \tau_{f} \sqrt{\rho}\xi(t) \\
    & \tau_{s} \frac{d S_{s}}{d t} = \omega \tau_{s}\rho -S_{s} \\
    & \tau_{i} \frac{d S_{i}}{d t} = \omega \tau_{i}\rho -S_{i}
    \end{aligned}\right.
\end{equation}
\par 
The dynamics of the closed system (equation~\ref{eq: closed_system}) at the critical point as shown in Fig~\ref{fig1}, are shown in Fig~\ref{fig6}. The error function in equation~\ref{eq: estimation} is the intrinsic nonlinear property that induces oscillation transition in the coarse-grained model. The firing rate corresponds to the numerical simulation of the spiking model, in which the network shows occasional bursting events and moderate synchronization. In the supercritical regime, these oscillations are indeed limit cycles in the $S_{s}-S_{i}$ plane, as shown in Fig~\ref{fig6} B, that collapse to a focus with decreasing fast-acting conductance. At approximately the critical parameter $\rm AMPA=2.0$, the limit cycle disappears, and a focus node occurs. The mechanism is explained by a Hopf bifurcation in the field equations through stability analysis. When $\rm AMPA$ increases, the fixed point will lose its stability through a supercritical Hopf bifurcation. Power-law criticality occurs at the bifurcation point and is influenced by external noise and network fluctuations. The remaining stable fixed point corresponds to a low and stable firing behavior for the network neurons. In the presence of noise, neurons occasionally cross the bifurcation or return back, causing neuronal avalanches on a small or large time scale. 

\subsection{Identification of the parameter of criticality}
Tracking the dynamics of the neuronal network with biophysical recordings is always a critical question in the field of computing modeling. Especially for critical neuronal activity, the network is extremely sensitive and its corresponding parameter is difficult to identify. In the parameter space with rich spiking patterns mentioned above, we aim to fit the network model to its corresponding functional MRI (BOLD) signal by estimating the appropriate synaptic weight. moving toward a Bayesian inference, the mathematical model is based on the neuronal network model and the Balloon-Windkessel model that takes the neural activity quantified by the spike rate of a pool of neurons and outputs the BOLD signal~\cite{friston2000nonlinear}. In the following, the synthetic data from the Balloon model are taken as the target biological signal to verify the effectiveness of our proposed method (detail in~\nameref{sec: Methods}).

\par 
The augmented Kalman filter is proven to be an efficient technique in parameter estimation in a combined deterministic-stochastic setting\cite{lourens2012augmented}. The unknown parameters are included in the state vector and estimated in conjunction with the states with a standard Kalman filter. In our problem, we hope to estimate a fixed parameter and enable real-time tracking of the nonstaining network. We use the ensemble version of the Kalman filter to implement sequential updating instead of reweighting in degeneracy problems~\cite{katzfuss2016understanding}. The augmented vector, consisting of all latent state variables and synaptic parameters, can be estimated in real-time using the Ensemble Kalman filter. After assimilation, we resimulate the network only by manipulating the synaptic weights with its assimilated parameters and compute the Pearson correlation between the simulated signal and target BOLD time series. 

\par
\begin{figure}
\centering	
\includegraphics[width=0.9\linewidth]{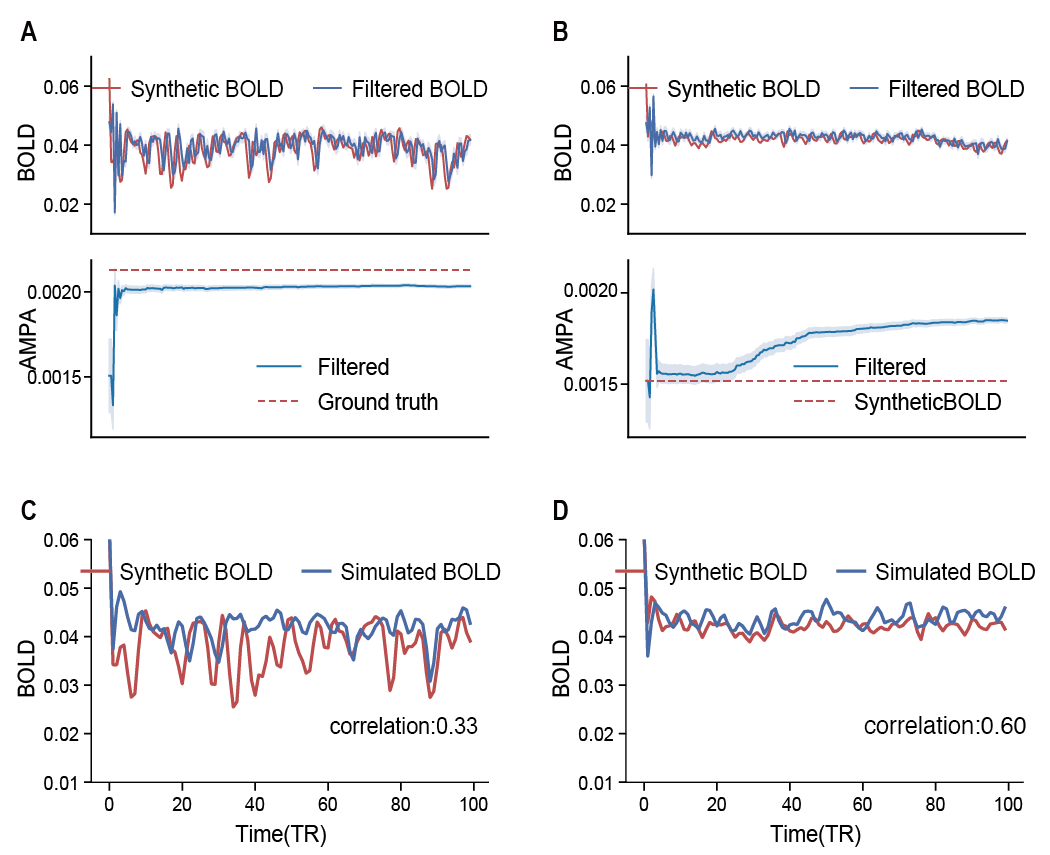}
\caption{{\bf Tracking the critical and subcritical dynamics in the network.} (A): The assimilation process of the network in tracking critical dynamics. The filtered signal is almost consistent with the ground truth(top panel). The filtered $\rm AMPA$ weights are plotted in the assimilation process (bottom panel). The dashed blue region represents the deviation among ensemble members. Note that the parameters converge rapidly to the ground truth of the critical parameters. (B): The same as (A) but in the subcritical state. (C) The ground truth and the resimulated BOLD signal show a correlation of 0.33 after a period of transition. The simulated signal tracks the critical BOLD signal in real time. (D) The simulated network reaches a correlation coefficient of 0.60 with its biological counterpart at the subcritical state.}
\label{fig7}
\end{figure}

Our proposed method is validated on synthetic data both in critical and subcritical networks. As before, we consider the two typical states in Fig~\ref{fig1} B and model them as the target networks. It is obvious that these two BOLD signals are different in amplitude: the critical BOLD fluctuates more strongly than the subcritical BOLD (\ref{fig7}A and B top panel). Collective firing in a critical state causes peaks in the firing rate, leading to a strong fluctuation in the BOLD signal. In contrast, the network firing remains stable around a fixed point in the subcritical state, and thus, the BOLD signal oscillates weakly. By using the proposed method, the parameter converges quickly to near the ground truth, while the filtered BOLD signal almost coincides with the synthetic BOLD signal (Fig~\ref{fig7}A bottom panel). In the subcritical state, the additive parameter constraint seems to be destructive, resulting in the parameters converging slowly to a wrong estimation. We use the estimated parameter to simulate the model and reach a correlation of 0.33 and 0.60 in these two situations respectively (Fig~\ref{fig7} C and D). Interestingly, the simulated subcritical network has a higher correlation than the critical one, although with much worse estimated parameters. This can be explained by the fact that the critical state is more sensitive to synaptic parameters than the subcritical state.

\par
By our proposed method, the synaptic weights are sequential and then converge to near the ground truth. The resimulation with estimated parameters demonstrates our enabling real-time tracking of nonstationary networks, including the critical network.

\section{Discussion}
Brain networks are proven to show rich dynamical patterns, called spontaneous activity, which do not look random and entirely noise-driven but are structured in spatiotemporal patterns~\cite{takeda2016estimating, vickers2017animal}. In the local network of the conductance-based neurons, we observe multiscale spiking features, including asynchronous irregular, synchronous regular and power-law neuronal avalanches. These findings seem to be meaningful because the similar co-emergence of rich cortical activities has been observed in both experimental and computational recordings~\cite{gireesh2008neuronal}. In particular, we show that fast-acting synapses modulate spiking behaviors in a wide dynamic range, and give rise to synchronous firing of neurons only in the mean-dominated regime. In contrast, the fast-acting synapse has little effect on the modulation of neuronal bursting events under a fluctuation-dominated background current. In this scenario, it appears that the influence of synaptic coupling is relatively weaker compared to the background current. The network tends to exhibit stationary firing activity and remains in focus in the dynamical regime. This suggests that the network dynamics are primarily driven by the background current rather than chemical synaptic interactions~\cite{vivekanandhan2023dynamical}.
\par 
Moreover, we found that an appropriate level of fast-acting synaptic coupling strength could trigger spike-based avalanches and evoke power-law avalanche criticality. In the context of our work here, criticality occurs in the boundary in the space of possible dynamical regimes. On the left side of the boundary, network neurons evolve independently of each other and fire occasionally, resulting in asynchronous population dynamics. On the other side, the population tends to work in the form of synchronous firing and synchronous terminating. At criticality, population dynamics are more diverse, occasionally oscillating and exhibiting moderate synchronous behaviors at large or small time scales. Different from the DP universality critical class, the model here reveals a power-law avalanche distribution with exponents $\tau_{s} = 1.6$ and $\tau_{t} = 2.1$. The network presents an oscillation with a peak frequency in the $\beta$ range and can be regulated with the time constant of the inhibitory synapse. These results highlight the importance of network coupling, including excitation and inhibition, in the evoking and modulating of rhythmic oscillation. 
\par 
Through a mean-field description of the neuronal network, we can predict the transition from asynchronous spiking to a sparse synchronous state through a Hopf bifurcation. Neuronal avalanches can be interpreted as a stochastic crossing behavior around the bifurcation point. A synchronous firing event refers to a phenomenon where a chain reaction of neuronal firing is evoked in a network. In other words, when a small subset of excitatory neurons spike, their collective excitation is sufficient to trigger a majority of the network's neurons to surpass their firing threshold and spike as well. In this model, we have presented three stages of neuronal synchronous oscillations: mean-dominated subthreshold dynamics, fast-initiating a spike event and time-delayed inhibitory cancellation. The degeneracy of population oscillation may evolve into a critical avalanche phenomenon in the presence of noise.
\par 
The identification of the parameter range for a critical network can be challenging due to its narrow and elusive nature. However, a potential approach to address this issue is by employing an Ensemble Kalman Filter (EnKF) in conjunction with the neuronal network and Balloon model, as proposed by Lu et al.~\cite{lu2022human} in their study on human brain networks. This technique offers a statistically driven estimation of network connectivity, specifically targeting the identification of the fast-acting coupling strength within a network exhibiting moderate nonstationarity, whether in critical or supercritical states. By applying the EnKF methodology, we were able to estimate a parameter that closely approximates the ground truth value, which serves as an accurate representation of the critical state. Indeed, applying the Ensemble Kalman Filter (EnKF) methodology to real data obtained from biological experiments poses certain challenges. One of the key challenges is ensuring that the observation signal aligns with the frequency and amplitude characteristics of the network behavior being studied. This problem deserves further exploration in our future modeling studies. Overall, the ability to estimate the critical parameter accurately contributes to our understanding of the network's behavior and its transition between different dynamical states.

\par 
In conclusion, our study provides a comprehensive exploration of coupled networks with conductance-based neurons and sheds light on the regulation of avalanche criticality. Our findings provide valuable insights for understanding and modeling criticality, offering implications for large-scale brain dynamics modeling and computational studies.

\section{Acknowledgements}
This work is jointly supported by the National Key R\&D Program of China (No.2019YFA0709502), the National Key R\&D Program of China (No.2018YFC1312904), the Shanghai Municipal Science and Technology Major Project (No.2018SHZDZX01), ZJ Lab, and Shanghai Center for Brain Science and Brain-Inspired Technology.

\bibliographystyle{elsarticle-harv}  

\end{document}